\documentclass[12pt,preprint]{aastex}

\usepackage{amsmath}

\slugcomment{KSUPT -- 10/1  \quad  March 2010}

\begin{document}

\title{Constraining dark energy with gamma-ray bursts}

\author{Lado Samushia\altaffilmark{1,2} and Bharat Ratra\altaffilmark{1}}

\altaffiltext{1}{Department of Physics, Kansas State University, 116 Cardwell
Hall, Manhattan, KS 66506, USA\\
lado@phys.ksu.edu, ratra@phys.ksu.edu.}
\altaffiltext{2}{National Abastumani Astrophysical Observatory, Ilia State
University, 2A Kazbegi Ave, GE-0160 Tbilisi, Georgia.}

\begin{abstract}

We use the measurement of gamma-ray burst (GRB) distances to
constrain dark energy cosmological model parameters. We employ two
methods for analyzing GRB data --- fitting luminosity relation of
GRBs in each cosmology and using distance measures
computed from binned GRB data. Current GRB data alone
cannot tightly constrain cosmological parameters and allow for a wide 
range of dark energy models. 

\end{abstract}


\section{Introduction}

The combination of recent measurements of distant supernovae Type Ia (SNe Ia)
apparent magnitudes \citep{sahni08, cunha09, perivolaropoulosshafieloo09,
hicken09}, cosmic microwave background anisotropy \citep{dunkley09, komatsu09},
the baryon acoustic signal in the power spectrum of galaxies
\citep{eisensteinetal, cole05, percivaletal07a, samushia09a}, and galaxy cluster
gas mass fractions \citep{allen08, samushia08, ettori09} indicates at high
confidence that about 70\% of the energy budget of the universe comes from
non-luminous dark energy that is close to spatially uniform and has negative
pressure.\footnote{This assumes that Einstein's general relativity provides an
accurate description of gravitation on cosmological scales. Attempts to do away
with dark energy by modifying general relativity are discussed by
\citet{hellwing09}, \citet{sen09}, \citet{shaposhnikov09}, \citet{setare09},
\citet{harko08}, \citet{capozziello08}, \citet{bamba09}, \citet{aluri09}, and
references therein. For a recent review see \citet{silvestri09}.}

Although the existence of dark energy is now a well-established observational
fact its physical nature is still a topic of great debate. The simplest and
historically first physical model of dark energy is the cosmological constant
$\Lambda$ that has an equation of state $p=-\rho$, where $p$ is its pressure and
$\rho$ the energy density \citep{pee84}. The cosmological constant is introduced
by hand into the equations of general relativity but could be related to the
energy of the fluctuating vacuum. This $\Lambda$CDM model, although simple and
in good accord with most available cosmological  data\footnote{The $\Lambda$CDM
model assumes the cold dark matter (CDM) model of structure formation which
might have some observational inconsistencies \citep[see, e.g.,][and references
therein]{peebles03, perivolaropoulos09, primack09}.} \citep[e.g.,][]{frieman08},
has a number of theoretical shortcomings \citep[e.g.,][]{ratra08}. These include
the so-called ``smallness'' and ``coincidence'' problems. The measured value of
the cosmological constant energy scale is of order $10^{-3}\rm eV$, some 30
orders of magnitude smaller than the Planck scale, perhaps what we would expect
the vacuum energy scale to be based on simple quantum field theoretical
arguments. In addition, since the energy density of nonrelativistic matter is
decreasing with the expansion of the Universe while the energy density of a
cosmological constant is constant, there is a very narrow window in time when
both of these have comparable energy densities, and it is unclear why we happen
to live at this special time.

Because of these and other issues a number of different dark energy models have
been considered. Typically these introduce a new component that acts like a
cosmological constant, in that it is close to spatially homogeneous, while
gradually decreasing in time to the small currently observed value. An early
example of this class of models is the $\phi$CDM model in which dark energy is
taken to be a scalar field $\phi$ \citep{peebles88, ratra88}. In this model, the
current cosmological constant energy scale is small because the Universe is old.
For recent reviews of dark energy see \citet{caldwell09}, \citet{frieman09}, and
\citet{sami09}.\footnote{In the $\phi$CDM model, we consider in this paper the
dark energy scalar field couples to the matter only through gravity. For dark
energy models with less restrictive couplings see \citet{wangzhang08},
\citet{dent09}, \citet{lavacca09}, \citet{jamil09}, \citet{chong09},
\citet{nesseris09}, and references therein. For other dark energy models see
\citet{bilic09}, \citet{basilakosperivolaropoulos08}, \citet{grande09},
\citet{dutta09}, \citet{feng09}, \citet{andrianov09}, and references therein.}

Although available cosmological data can constrain dark energy models, they are
not yet good enough to strongly discriminate between different dark energy
models (see, e.g., \citealt{gong09}, \citealt{kilbinger09}, \citealt{coc09}, and
references therein). In the near future better-quality and more independent
cosmological observations should be able to break this degeneracy \citep[see,
e.g.,][]{wang09, mortonson09, thomas09, arun09, yashar09}.\footnote{In the near
future, measurements of nonlinear structure formation will help discriminate
between different dark energy models, see, e.g., \citet{grossi09},
\citet{francis09}, \citet{casarini09}, and references therein. Other tests that
hold significant potential are the angular size of radio sources and galaxies as
a function of redshift \citep[e.g.,][and references therein]{daly09, santos08}
and measurement of the Hubble parameter as a function of redshift
\citep[e.g.,][and references therein]{samushia06, lin08, dev08,
fernandezmartinez08}.}

SNe Ia were one of the first cosmological probes to give direct evidence for dark
energy. These are very bright exploding stars and are standardizable candles
that can be seen to very large distances. More then 300 well-calibrated distant
SNe Ia have already been observed up to the redshift of 1.6 \citep{kowalski09,
hicken09}. These measurements alone give more then 5$\sigma$ evidence for the
existence of dark energy, but they are not very effective in constraining dark
energy model parameters overall unless used in combination with other data. Even
when current SNe Ia data are combined with all other currently available data
sets, we cannot yet determine if the energy density of dark energy is constant
as required by the $\Lambda$CDM model or if it varies in time as suggested by
dynamical alternatives such as $\phi$CDM. A dedicated SNe Ia space mission should
result in significantly more higher quality data that should help resolve this
issue \citep[see, e.g.,][and references therein.]{podariu01a, alam09}.

One way of improving our understanding of how dark energy behaves is to study
the evolution of the universe at redshifts higher than those probed by SNe Ia.
This requires standard candles that are visible at great distances. Gamma-ray
bursts (GRBs) could in principle serve as such high redshift standardizable
candles. They are the most luminous events in the universe today and can be seen
beyond $z=8$ (see, e.g., \citealt{tanvir09}).\footnote{For a review of GRB
physics see, e.g., \citet{meszaros06}.} If it is definitely established that
GRBs are standardizable candles, their visibility at high-redshift should prove
to be very useful in discriminating between $\Lambda$CDM and time-varying dark
energy models.\footnote{For early discussions of the use of GRBs as a cosmology
probe see, e.g., \citet{lamb00}, \citet{nemiroff00}, \citet{ghirlanda04},
\citet{friedman05}, \citet{firmani05}, \citet{xu05}, \citet{mortsell05},
\citet{digirolamo05}, \citet{bertolumi06}, and \citet{lamb05}. More recent
studies may be traced back through \citet{mosquera08}, \citet{amati08},
\citet{basilakosperivolaropoulos08}, \citet{capozzielloizzo08},
\citet{tsutsui09}, and \citet{qi09}.}

With the intention of getting cosmological constraints from GRB observations a
number of GRB calibrations  have been used so far \citep[see, e.g.,][and
references therein]{schaefer07}. One that gives least scatter and therefore most
information is 
\begin{equation}
\label{EE}
\log\left(\frac{E_{\gamma}}{1\ \rm erg}\right) = A_1 +  B_1 \log\left(\frac{E_{\rm peak}(1+z)}{300\ \rm keV}\right),
\end{equation}
\noindent
a relation that connects the total burst energy of the GRB ($E_{\gamma}$) to the
peak energy of the GRB spectrum ($E_{\rm peak}$) \citep{ghirlanda04}.
Regrettably, we do not yet have a model-independent way of computing the
coefficients $A_1$ and $B_1$. A better understanding of physical processes that
result in the burst, or observations of nearby GRBs (to which distances can be
measured independently), could in principle help us to calibrate the
$E_{\gamma}$-$E_{\rm peak}$ relation without any prior assumptions. To extract
cosmological information, GRBs have to be recalibrated for every dark energy
model considered (at each set of parameter values).  This is time consuming and
also results in large statistical uncertainties and hence GRB cosmological
constraints that are poor. 

Recently, methods of calibrating GRBs in cosmology-independent manners have been
proposed and used to constrain some dark energy models \citep[see e.g.,][that
use SNe Ia measurements to externally calibrated GRBs]{kodama08, liangetal08,
wei08, liang08}.  The resulting cosmological constraints are still loose, but in
the future when more high precision GRB observations become available this could
provide a strong test of dark energy.

\citet{yuwang08} recently used data of 69 GRBs \citep{schaefer07} to construct a
distance measure that can be used to constrain cosmological models. The
advantage of this method is that internally calibrated GRB data may be
straightforwardly combined with other data when deriving cosmological
constraints. On the other hand, with this method the resulting cosmological
results are sensitive to the chosen binning. This method also requires an input
cosmological model and thus is not completely cosmology independent.  When this
method is used to constrain $\Lambda$CDM the GRB data favor lower values of both
cosmological constant energy density ($\Omega_\Lambda$) and nonrelativistic
matter energy density ($\Omega_{\rm m}$) than do the SNe Ia data. The GRB data by
themselves are unable to strongly constrain cosmological parameters, for example
in spatially flat $\Lambda$CDM the GRB data require $\Omega_{\rm
m}=0.25^{+0.12}_{-0.11}$ at 1$\sigma$ confidence \citep{yuwang08}.

In this paper we use GRB data to constrain time variation of dark energy's
energy density. First we recalibrate GRBs for each cosmological model and
compare the result with the ones derived using the data and method of
\citet{yuwang08}. We consider a $\phi$CDM model where a scalar field which is
close to spatially uniform on cosmological scales slowly rolls down an almost
flat potential and plays the role of dark energy. 
 
In the next section, we summarize the dynamics of this scalar field dark energy
model. In Section\ 3, we describe our methods and computations. We present and
discuss our results in Section\ 4.

\section{Scalar field as dark energy}

In the $\phi$CDM model, consistent with the indications from cosmic microwave
background anisotropy
measurements \citep[e.g.,][]{podariu01b,page03}, we only consider the
spatially flat universe case. The invariant four-interval in a homogeneous and
isotropic version of such a universe is 
\begin{equation}
\label{ds}
ds^2=-dt^2+a^2(t)d\vec{x}\cdot d\vec{x},
\end{equation}
\noindent
where $t$ is cosmic time, $d\vec{x}$ is the spatial separation in three-dimensional 
Euclidean space and $a(t)$ is the time-dependent scale factor which
determines the change in distance between two distant noninteracting test
particles in the universe.

In the set of models we consider the scalar field $\phi$ with Lagrangian density
\begin{equation}
\label{lagrangian}
\mathcal{L}=\frac{1}{2}\partial_\mu\phi\partial^\mu\phi-\frac{1}{2}V(\phi)
\end{equation}
\noindent
is the dark energy. Here $V(\phi)$ is the potential energy density of the scalar
field.  In the expanding, spatially homogeneous and isotropic universe described
in Equation\ \eqref{ds} the spatially homogeneous scalar field obeys the modified
Klein-Gordon equation,

\begin{equation}
\label{phi}
\ddot\phi + 3\frac{\dot a}{a}\dot\phi + \frac{1}{2}\frac{\partial V(\phi)}{\partial \phi} = 0,
\end{equation}
\noindent
and the dynamics of the scale factor is governed by 

\begin{equation}
\label{friedman}
\left(\frac{\dot a}{a}\right)^2=\frac{8\pi G}{3}(\rho_{\rm m} + \rho_\phi).
\end{equation}
\noindent
In Equation\ (\ref{friedman}), $G$ is the universal gravitational constant, and
$\rho_{\rm m}$ and $\rho_\phi$ are the energy densities of nonrelativistic
matter and the scalar field, respectively. If the scalar field is uniform in
space then, from Equation\ (\ref{lagrangian}), the energy density of the scalar field
is

\begin{equation}
\rho_\phi=\frac{1}{32\pi G}\left(\dot\phi^2+V(\phi)\right).
\end{equation}

Since an underlying more fundamental explanation of dark energy remains elusive,
there is as yet no first principles way of choosing the scalar field potential.
If we choose the potential energy density to be inversely proportional to a
power of the scalar field, $V(\phi)\sim \phi^{-\alpha}$, the $\phi$CDM model has
a number of very interesting features \citep{peebles88, ratra88}. First, even if
the scalar field starts off from a very high energy density state, its energy
density decreases to a very small value during the course of cosmic evolution.
Second, in the radiation and matter-dominated epochs the evolution of the scalar
field ``tracks'' the evolution of the dominant component. The scalar field
slowly comes to dominate, leading to the end of matter domination and the start
of the scalar field-dominated epoch. So, in the $\phi$CDM scenario the
``smallness'' and ``coincidence'' problems mentioned above are partially
resolved because of the time-evolution properties of the scalar field. Even if
the scalar field potential is not exactly inverse power law, this form of
potential provides a very economic way of parameterizing the slowly-evolving
dark energy scenario with just one positive parameter $\alpha$. Moreover, unlike
the XCDM parameterization of dark energy, the $\phi$CDM model is physically
consistent \citep[see, e.g.,][]{ratra91}.

In the $\phi$CDM model, the dark energy density, unlike the cosmological
constant, varies slowly in time. The dark energy density increases as we go back
in time and larger values of $\alpha$ correspond to faster evolution of dark
energy. As a result, observable quantities such as luminosity and angular
diameter distances in the $\phi$CDM model differ from the predictions of the
$\Lambda$CDM model with a time-independent cosmological constant. Some
predictions, of course, depend also on the values of other cosmological
parameters, such as the Hubble constant $H_0$ or the energy density of baryonic
matter $\Omega_{\rm b}$, and so many independent cosmological tests are required
to break this degeneracy and constrain dark energy model parameters.

A number of cosmological tests have been used to constrain the $\phi$CDM model,
including the angular sizes of radio sources and quasars as a function of
redshift \citep{chen03a, podariu03, daly09}, the apparent magnitude of SNe Ia as
a function of redshift \citep{wilson06, samushia09b}, and the gas mass fraction
of large relaxed clusters as a function of redshift \citep{chen04, samushia08}.
Current cosmological observations are in good agreement with a time-independent
cosmological constant in a close to spatially flat $\Lambda$CDM model, but
slowly rolling scalar field dark energy in the $\phi$CDM model is not yet ruled
out at high confidence. 

In this paper we use measured GRB luminosity distance as a function of redshift
to constrain slowly rolling scalar field and other dark energy models. Because
of the calibration problems mentioned above, GRB data alone cannot constrain
cosmological parameters effectively. We also combine the results obtained from
the GRB analysis with constraints from SNe Ia and baryon acoustic peak
measurements, to illustrate the effect and weight of current GRB data in such a
combined analysis. 

\section{Cosmological constraints from GRB}

Besides the $E_{\rm peak}$--$E_\gamma$ relation, (Equation\ \eqref{EE}),
\citet{schaefer07} uses four other calibrations for GRBs that relate total
luminosity to directly measurable quantities. These calibration relations are
given by
\begin{eqnarray}
\log\left(\frac{L}{1 \rm erg s^{-1}}\right) &=& A_2 + B_2 \log\left(\frac{\tau_{\rm lag}(1+z)^{-1}}{0.1\ \rm s}\right),\nonumber\\
\log\left(\frac{L}{1 \rm erg s^{-1}}\right) &=& A_3 + B_3 \log\left(\frac{V(1+z)}{0.02}\right),\nonumber\\
\label{eq:calibr}
\log\left(\frac{L}{1 \rm erg s^{-1}}\right) &=& A_4 + B_4 \log\left(\frac{E_{\rm peak}(1+z)}{300 \rm KeV}\right),\\
\log\left(\frac{L}{1 \rm erg s^{-1}}\right) &=& A_5 + B_5 \log\left(\frac{\tau_{\rm RT}(1+z)^{-1}}{0.01 \rm s}\right)\nonumber,
\end{eqnarray}
where $L$ is the absolute luminosity, $\tau_{\rm lag}$ is the GRB lag time (the
time shift between the hard and soft curves), $V$ is the variability (the
normalized variance of an observed light curve around the smoothed light curve),
$E_{\rm peak}$ is the peak energy of the GRB, and $\tau_{\rm RT}$ is the rise
time or the time over which the light curve rises by half of the peak flux. 

Following \citet{schaefer07} we take Equations~(\ref{eq:calibr}) and (\ref{EE}) and
for each cosmological model find the best-fit values for the $A$ and $B$
parameters using the bisector least-square method \citep[for a description
see][]{isobe90}. We use the best-fit $A$ and $B$ values and the measured
$\tau_{\rm lag}$, $V$, $E_{\rm peak}$, $\tau_{\rm RT}$ values to compute $L$ and
$E_\gamma$ using the same Equations~(\ref{eq:calibr}) and (\ref{EE}). For each
calibration relation we then compute the luminosity distance as
\begin{eqnarray}
\label{eq:ld}
d_{\rm L}^2 &=& \frac{L}{4\pi P_{\rm bolo}},\nonumber\\
d_{\rm L}^2 &=& \frac{E_\gamma(1+z)}{4\pi S_{\rm bolo}F_{\rm beam}},
\end{eqnarray}
where $P_{\rm bolo}$ is the bolometric peak flux and $S_{\rm bolo}$ is the
bolometric fluence of the GRB. $F_{\rm beam}$ is so-called beam factor $F_{\rm beam} = 1
- \cos{\theta}_{\rm jet}$, where $\theta_{\rm jet}$ is a jet opening angle.

We then derive an effective luminosity distance by weighting the five 
estimates of GRB luminosity distances,
\begin{eqnarray}
\log(\bar{d}_L^2(z_i)) = \frac{\sum_\alpha{\log(\bar{d}_L^2(z_i)_\alpha)}/\sigma^2_{i,\alpha}}{\sum_\alpha{1/\sigma^2_{i,\alpha}}},\\
\sigma^2(\log({\bar{d}_L^2(z_i)})) = 1/\sum_\alpha{1/\sigma^2_{i,\alpha}},
\end{eqnarray}
where index $i$ runs over 69 redshift bins, and $\alpha$ over five
calibration relations.
To constrain cosmological parameters we use $\chi^2$ defined by
\begin{equation}
\chi^2 = \displaystyle\sum_{\rm i = 1}^{69}\frac{(\log(\bar{d}_L^2(z_i))^{\rm obs}-\log(\bar{d}_L^2(z_i))^{\rm th})^2}{\sigma^2(\log(\bar{d}_L^2(z_i))}.
\end{equation}
\noindent

We also adopt the method of \citet{yuwang08} for using GRB data to constrain
cosmological parameters. She placed each of the 69 GRBs in the redshift range
$z=0.17$ to $z=6.6$ \citep{schaefer07} at a luminosity distance that minimized a
combined $\chi^2$ that took weighted account of all five calibration relations.
She then computed a distance measure
\begin{equation}
\label{dm}
\bar r_{\rm p}=\frac{r_{\rm p}(z)}{r_{\rm p}(0.17)},
\end{equation}
where 
\begin{equation}
\label{rp}
r_{\rm p}(z)=\frac{H_0}{hc}\frac{1}{z(1+z)^{1/2}}d_L(z),
\end{equation}
\noindent
and $d_L(z)$ is the luminosity distance at redshift $z$, $h=H_0/(\rm 100\ km\
s^{-1}\ Mpc^{-1})$, and $c$ is the speed of light. The ratio in Equation\ (\ref{dm})
does not depend on the Hubble constant and does not require information about
the absolute calibration of GRBs (which are unknown). 

\citet{yuwang08} computed the  distance measure $\bar r_{\rm p}$ in six redshift
bins $\bar r_{\rm p}(z_{\rm i})$, $i=1,2,\dots6$. The values of $\bar r_{\rm
p}(z_{\rm i})$ are shown in Table\ II and the normalized covariance matrix is
shown in Table\ III of \citet{yuwang08}. For currently viable cosmological
models, these $\bar r_{\rm p}(z_{\rm i})$ are almost completely independent of
the cosmological model and so provide a useful summary of current GRB data
\citep{yuwang08}. This information can be used to constrain any dark energy
model and the resulting GRB data constraints can be straightforwardly combined
with other constraints. In this approach, $\chi^2$ is given by
\begin{equation}
\chi^2(\Omega_{\rm m}, p) = \Delta(z_{\rm i}) \sigma_{\rm i} (S^{-1})_{\rm i j} \sigma_{\rm j} \Delta(z_{\rm j}),
\end{equation}
where
\begin{equation}
\Delta(z_{\rm i}) = \bar r_{\rm p}^{\rm data}(z_{\rm i}) - \bar r_{\rm p}^{\rm theory}(z_{\rm i}),
\end{equation}
$S_{\rm i j}$ is the normalized covariance matrix given in Table\ III of
\citet{yuwang08} and summation over repeated indexes is assumed. Here,
$\sigma_{\rm i}$ is $\sigma_{\rm i}^+$ if $\Delta(z_{\rm i})>0$ and $\sigma_{\rm
i}^-$ if $\Delta(z_{\rm i})<0$.

In this paper we consider three cosmological models, $\Lambda$CDM, the XCDM
parameterization of the dark energy equation of state $p_{\rm x}=\omega_{\rm
x}\rho_{\rm x}$ in a spatially flat universe, and the spatially flat $\phi$CDM
model. Since we are comparing low-redshift predictions to observations we ignore
the contribution of radiation in these models. In this case, in all three
models, the background evolution can be fully described by two parameters, the
fractional energy density of nonrelativistic matter $\Omega_{\rm m}$ and a
parameter $p$ that describes the properties of dark energy. In $\Lambda$CDM $p$
is the fractional energy density of the cosmological constant $\Omega_\Lambda$,
in XCDM it is the equation of state parameter $\omega_{\rm x}$, and in $\phi$CDM
it is the positive parameter $\alpha$ that governs the steepness of the scalar
field potential energy density. 

In each model, for both methods, we divide the two-dimensional space of
cosmological parameters into an equidistant grid, and for each pair of
parameters $\Omega_{\rm m}$ and $p$ we compute the theoretical luminosity
distance. We then compute the difference between the theoretical prediction and
the measured value at each of the 69 redshifts listed in Table 2 of
\citet{schaefer07} or the six redshifts listed in Table\ II of \citet{yuwang08},

The best-fit parameters are defined as the pair $(\Omega_{\rm m}^*, p^*)$ that
gives the minimum value of $\chi^2(\Omega_{\rm m}, p)$. The 1$\sigma$, 2$\sigma$, and 3$\sigma$
confidence level contours are defined as the sets of points where the value of
$\chi^2(\Omega_{\rm m}, p)$ is more than its minimum value $\chi^2(\Omega_{\rm
m}^*, p^*)$ by 2.30, 6.18, and 11.83, respectively. If the likelihood function,
$\propto\exp{(-\chi^2/2)}$, was Gaussian, the 3$\sigma$ contour would be 99.7\%
likely to enclose the true values of cosmological parameters. In our case the
likelihood function has a single maximum and decreases monotonically from the
best-fit value point, so the true values of cosmological parameters are very
unlikely to be outside the 3$\sigma$ contours we compute.

Figures 1--6 show the constraints from the \citet{yuwang08} GRB data on
$\Lambda$CDM, XCDM, and $\phi$CDM model parameters. Figures 7--12 show the
constraints derived using a method similar to that of \citet{schaefer07}.

\section{Results and discussion}

The GRB constraints shown in Figures\ 1--3 are consistent with the ``standard''
spatially flat $\Omega_\Lambda=0.7$ $\Lambda$CDM model at a little under
2$\sigma$, with the GRBs mildly favoring a somewhat lower value of $\Omega_{\rm
m}$ than the ``standard'' value  that is compensated by space curvature and
$\Omega_\Lambda=0$ (Figure\ 1) or a mildly time-varying dark energy (Figures\ 2 and
3). While the constraints of Figs.\ 1--3 derived from the \citet{yuwang08} data
have not previously been shown, these results are implicit in the discussions of
\citet{yuwang08} and other analyses of GRB constraints, including
\citet{kodama08}, \citet{liang08}, \citet{amati08}, and \citet{tsutsui09}.

These results, and the fact that the current GRB data contours are quite broad,
are probably an indication of the preliminary nature of the current GRB data.
Current GRB data by themselves are unable to effectively constrain cosmological
parameters. The GRB constraints however, are a little tighter than radio galaxy
angular size versus redshift constraints \citep[e.g.,][]{daly09}, constraints
from strong gravitational lensing data \citep[e.g.,][]{chae04}, and those from
the measurement of the Hubble parameter as a function of redshift
\citep[e.g.,][]{samushia07}.

To get tighter constraints on cosmological parameters, and to see how current
GRB data affect constraints derived from other data sets, we combine the results
of our GRB analysis with SNe Ia apparent magnitude versus redshift data
\citep[the Union data of][]{kowalski09} and measurements of the baryon acoustic
(BAO)
peak \citep{percivaletal07a}. Since all three sets of measurements are
independent, we define the total likelihood function of cosmological parameters
for the combined data as a product of the individual likelihood functions,

\begin{equation}
\label{eq:totallik}
\mathcal{L}_{\rm tot}=\mathcal{L}_{\rm GRB}\mathcal{L}_{\rm SN}\mathcal{L}_{\rm BAO},
\end{equation}
\noindent
and compute the best-fit values and confidence level contours from
$\mathcal{L}_{\rm tot}$ as before. The SNe Ia likelihood function
$\mathcal{L}_{\rm SN}$ depends on the assumptions that we make about the value
of the Hubble constant. Here we marginalize over $h=0.73\pm0.03$ with the

4--6 show the constraints on cosmological parameters of the $\Lambda$CDM, XCDM,
and $\phi$CDM models from a joint analysis of the SNe Ia Union and baryon
acoustic peak measurements, without and with the GRB data. These plots show that
current GRB data only marginally affect the joint SNeIa and BAO peak constraints
(which are amongst the tightest provided by current data), favoring slightly
lower values of the nonrelativistic matter density parameter $\Omega_{\rm m}$.

Figures 7--9 are similar to Figures\ 1--3, but derived by recalibrating GRB data
for each cosmology, using a method similar to that of \citet{schaefer07}.  The
``standard'' spatially flat $\Omega_\Lambda = 0.7$ $\Lambda$CDM model is about
2.5$\sigma$ from the best-fit value. In all three models GRB data favor a
nonrelativistic matter-dominated universe. Figures 10--12 show the joint
constraints from GRB, SNe Ia and BAO data. The constraints are dominated by the
SNe Ia and BAO data, with the GRB data shifting the best-fit values to slightly
larger values of $\Omega_{\rm m}$. 

The GRB constraints derived using the two different methods disagree with each
other at more than 2$\sigma$ confidence level (compare Figures\ 1--3 and 7--9).
This is somewhat worrying but not completely unexpected as the field is still
under rapid development.

GRB data alone do not provide tight constraints on cosmological parameters.
Moreover, while not greatly significant, current GRB data favor cosmological
parameter values that are at odds with what other data favor. When used in
combination with some of the highest-quality current data (e.g., SNe Ia and BAO
peak measurements) current GRB data only slightly change the results. This is
mainly because in the absence of an absolute calibration of GRBs they, as
standard candles, have big measurement uncertainties. This is however quite
likely to change as more and better-quality GRB measurements become available
with improvements in methods to calibrate GRBs. GRBs could potentially
provide a very strong test of the time variation of dark energy as they can be
observed up to redshifts beyond eight, at distances where other standard candles
cannot be detected. 

\acknowledgements
We thank the referee for a detailed and helpful report. We acknowledge 
support from DOE grant DE-FG03-99EP41093 and the Georgian National Science
Foundation grant ST08/4-442.

\begin{figure}
\includegraphics[clip, trim=10mm 0 10mm 0, width=180mm, height=180mm]{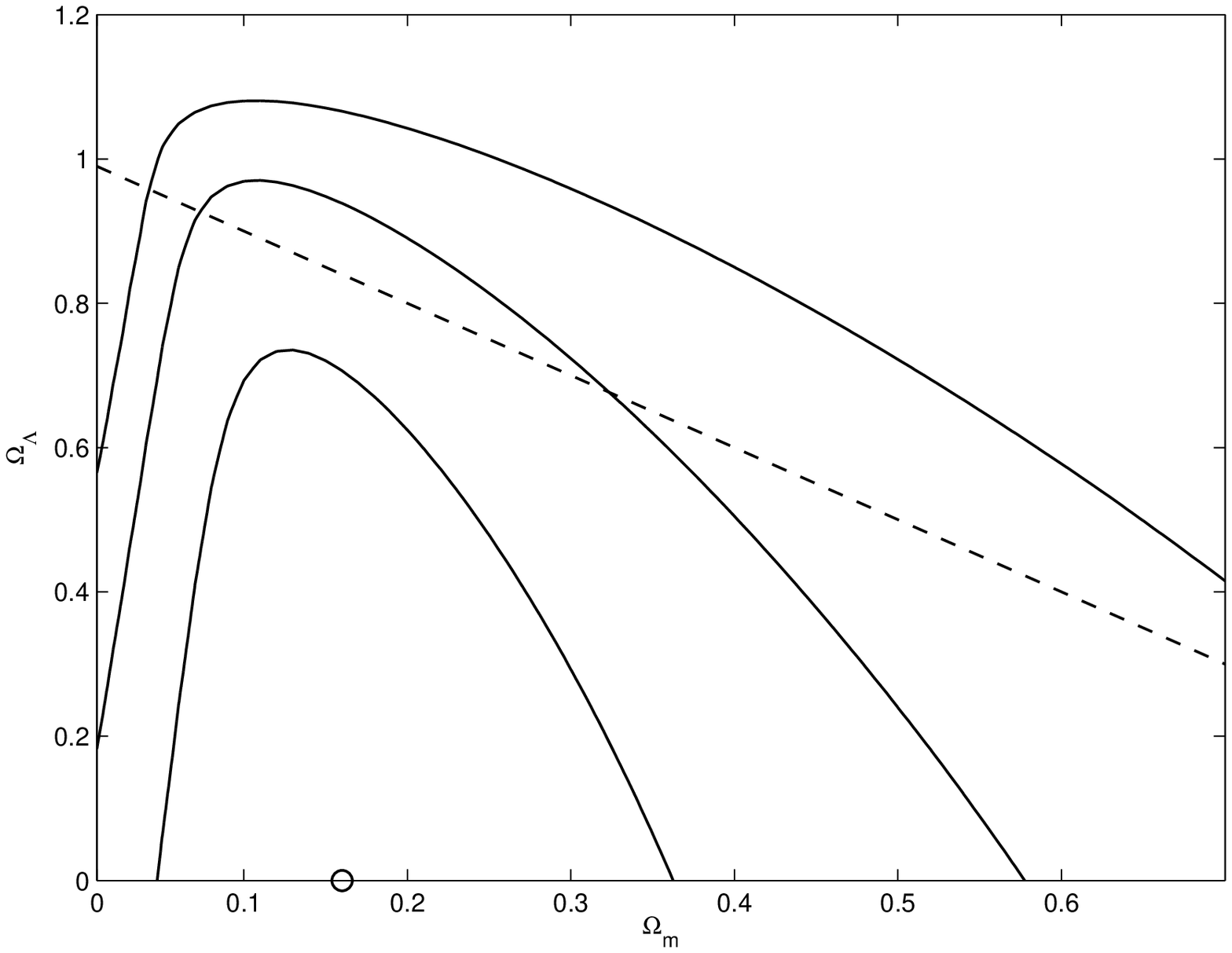}
\caption{1$\sigma$, 2$\sigma$, and 3$\sigma$ confidence level contours for the $\Lambda$CDM 
model from the GRB data, derived using the method of \citet{yuwang08}. The circle indicates best-fit parameter values $\Omega_{\rm m}=0.16$, $\Omega_\Lambda=0.0$ with $\chi^2=0.41$ for $4$ degrees of freedom. The dashed line demarcates spatially flat $\Lambda$CDM models.}
\end{figure}

\begin{figure}
\includegraphics[clip, trim=10mm 0 10mm 0, width=180mm, height=180mm]{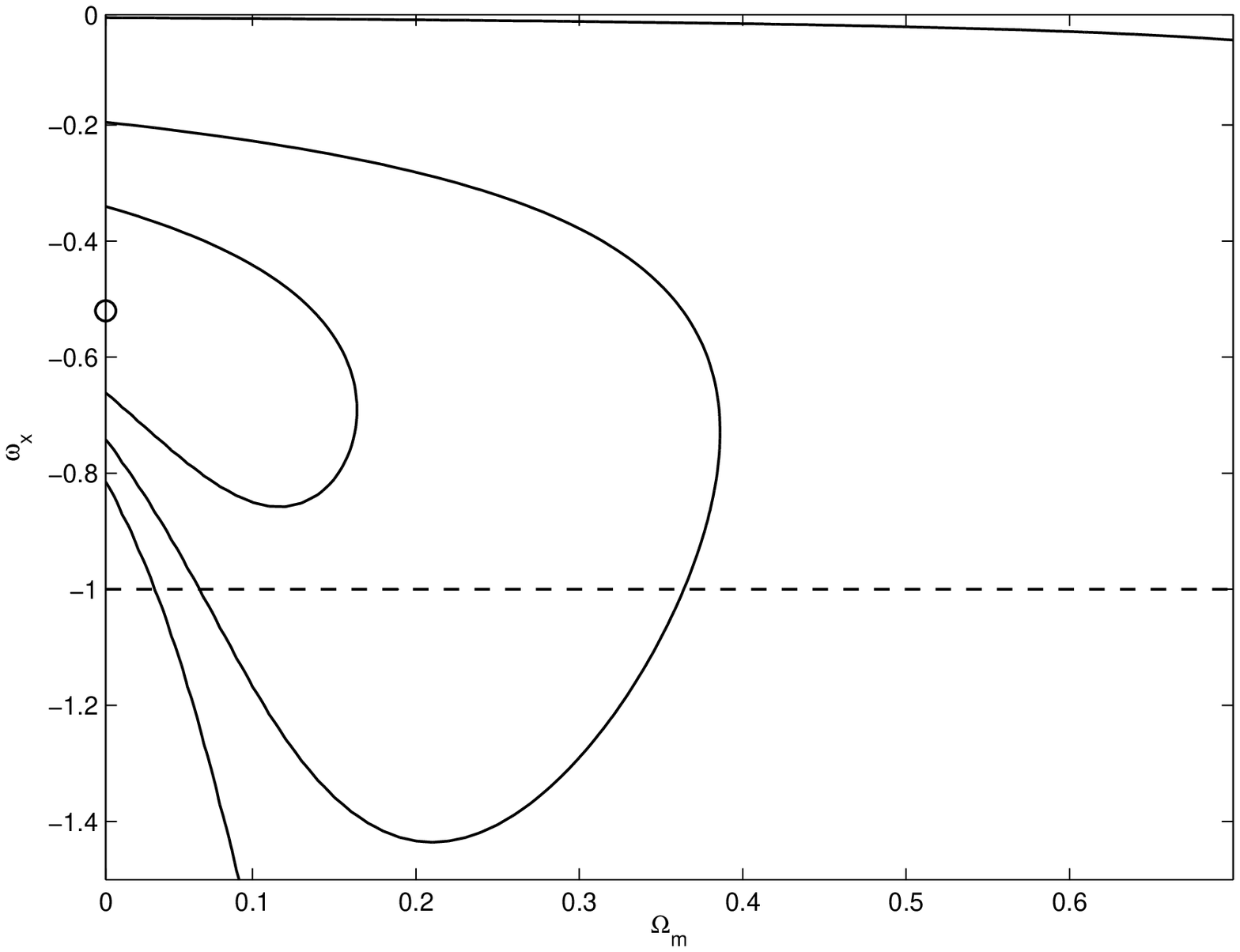}
\caption{1$\sigma$, 2$\sigma$, and 3$\sigma$ confidence level contours for the XCDM model from 
the GRB data, derived using the method of \citet{yuwang08}. The circle indicates best-fit parameter values $\Omega_{\rm m}=0.0$, $\omega_{\rm x}=-0.52$ with $\chi^2=2.17$ for $4$ degrees of freedom. The dashed line demarcates spatially flat $\Lambda$CDM models.}
\end{figure}

\begin{figure}
\includegraphics[clip, trim=10mm 0 10mm 0, width=180mm, height=180mm]{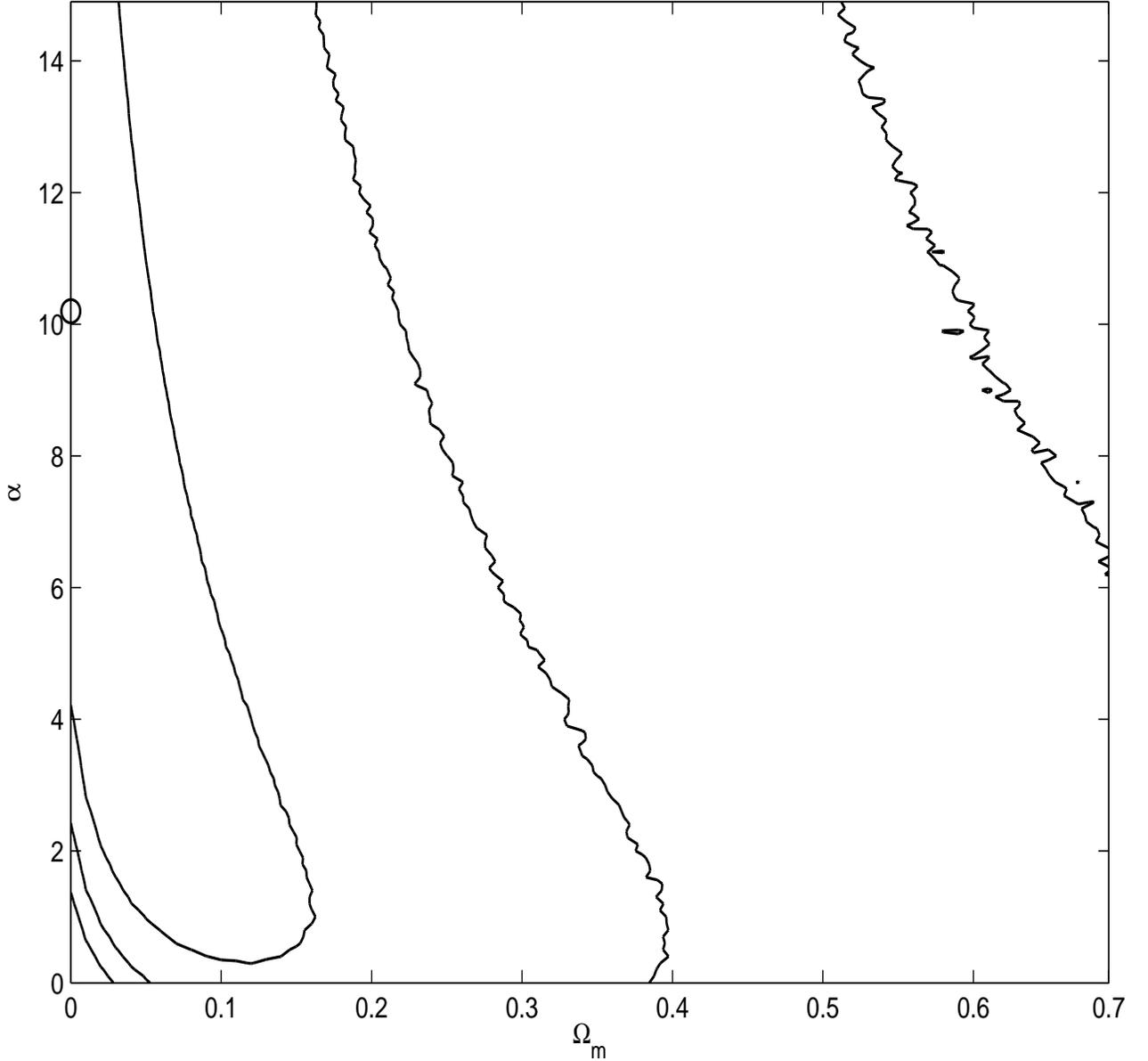}
\caption{1$\sigma$, 2$\sigma$, and 3$\sigma$ confidence level contours for the $\phi$CDM model 
from the GRB data, derived using the method of \citet{yuwang08}. Numerical noise is responsible for the jaggedness of some parts of contours. The circle indicates best-fit parameter values $\Omega_{\rm m}=0.0$, $\alpha=10.2$ with $\chi^2=1.39$ for $4$ degrees of freedom. The $\alpha=0$ horizontal axis corresponds to the spatially flat $\Lambda$CDM case.}
\end{figure}

\begin{figure}
\includegraphics[clip, trim=10mm 0 10mm 0, width=180mm, height=180mm]{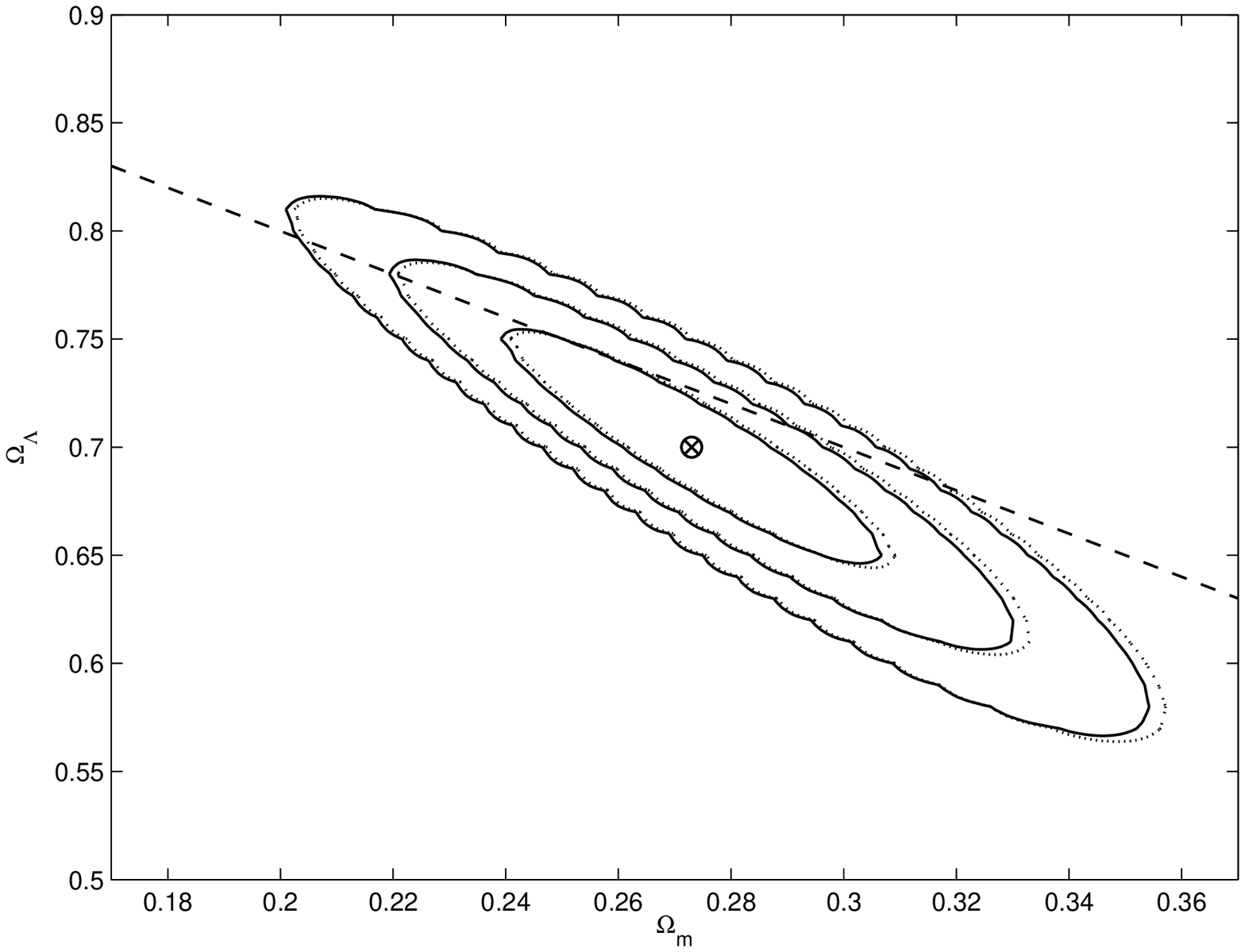}
\caption{1$\sigma$, 2$\sigma$, and 3$\sigma$ confidence level contours for the $\Lambda$CDM 
model (dashed line demarcates spatially flat models). Solid lines (circle 
denotes the best-fit point) are derived using the GRB data (method of 
\citep{yuwang08}), SNe Ia Union data, and BAO peak measurements, while dotted
lines (cross denotes the best-fit point) are derived using SNeIa and BAO data
only. Numerical noise is responsible for the jaggedness of some parts of
contours. The best-fit parameters in both cases are $\Omega_{\rm m}=0.27$,
$\Omega_\Lambda=0.7$ with $\chi^2=321$ for $307$ degrees of freedom (dotted
lines) and $\chi^2=326$ for $313$ degrees of freedom (solid lines). Note the
different axes scales compared to Figure\ 1.}
\end{figure}

\begin{figure}
\includegraphics[clip, trim=10mm 0 10mm 0, width=180mm, height=180mm]{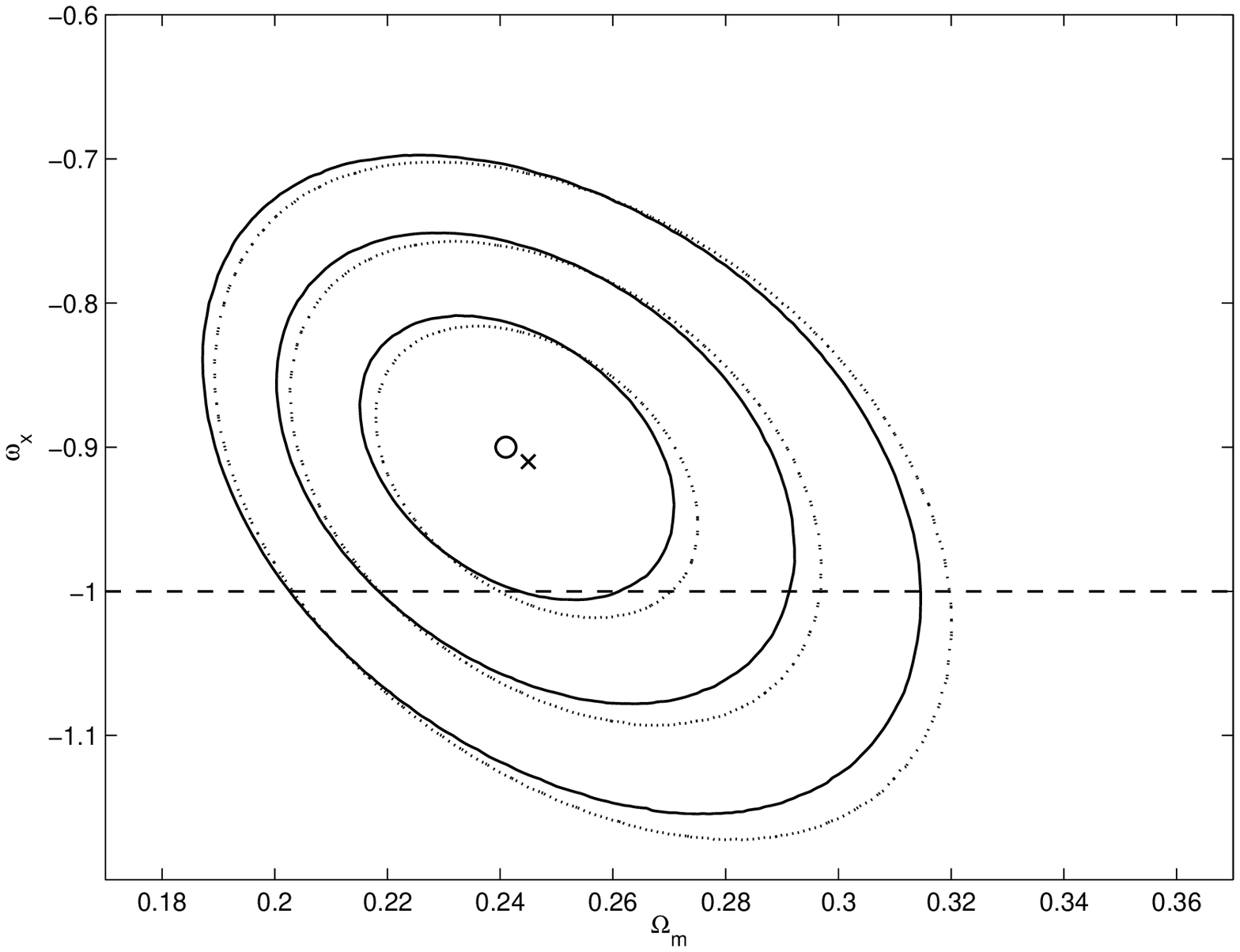}
\caption{1$\sigma$, 2$\sigma$, and 3$\sigma$ confidence level contours for the XCDM model. 
Solid lines (circle denotes the best-fit point) are derived using GRB data 
(method of \citep{yuwang08}), SNe Ia Union data, and BAO peak measurements, while
dotted lines (cross denotes the best-fit point) are derived using only SNe Ia and
BAO peak data only. The best-fit parameter values are: for solid contours
(circle)  -- $\Omega_{\rm m}=0.24$, $\omega_{\rm x}=-0.90$ with $\chi^2=327$ for
$313$ degrees of freedom, and for dotted contours (cross) -- $\Omega_{\rm
m}=0.25$, $\omega_{\rm x}=-0.91$ with $\chi^2=322$ for $307$ degrees of freedom.
Note the different axex scales compared to Figure\ 2.}
\end{figure}

\begin{figure}
\includegraphics[clip, trim=10mm 0 10mm 0, width=180mm, height=180mm]{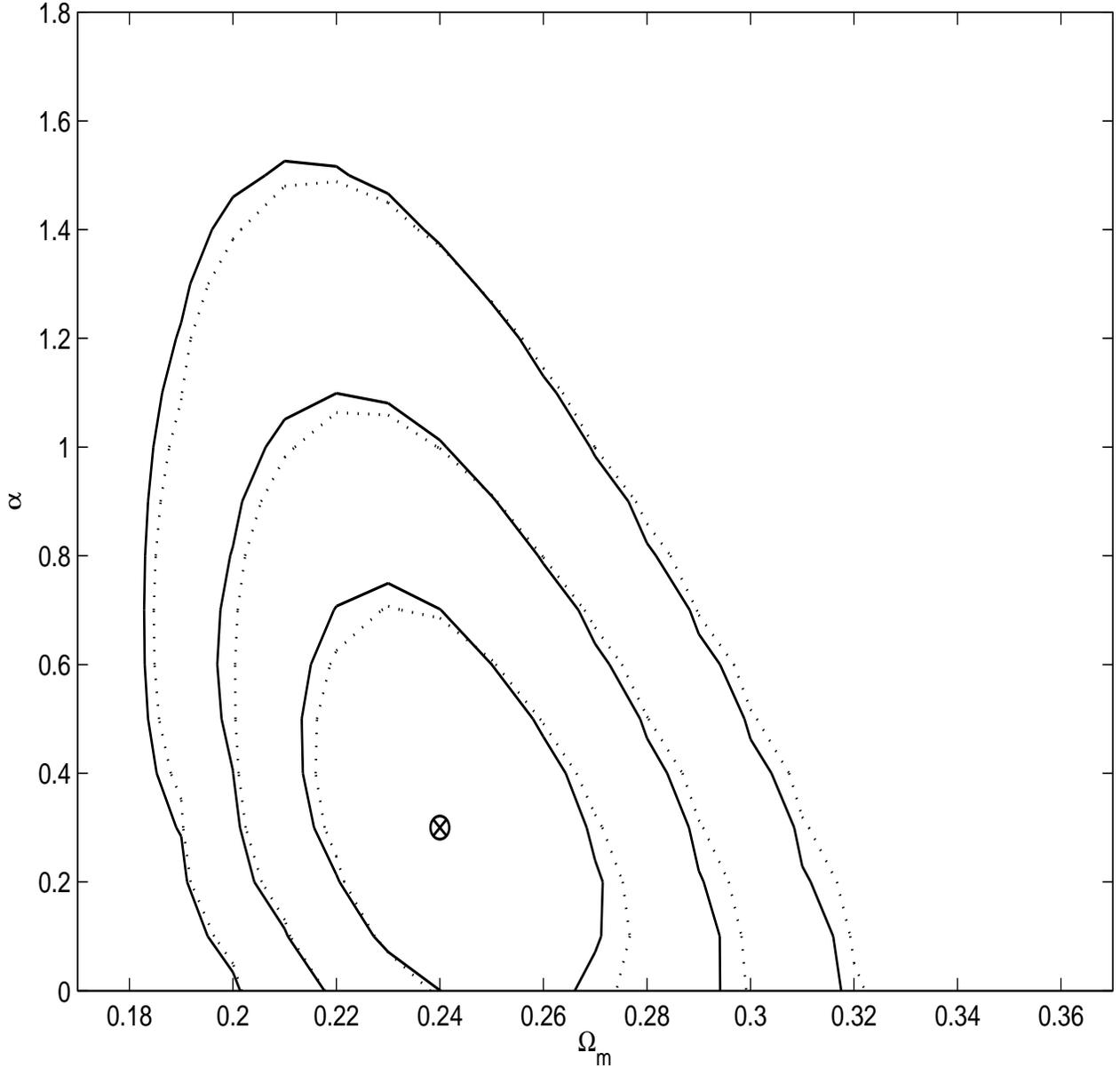}
\caption{1$\sigma$, 2$\sigma$, and 3$\sigma$ confidence level contours for the $\phi$CDM model.
Solid lines (circle denotes best-fit point) are derived using GRB data (method 
of \citep{yuwang08}), SNe Ia Union data, and BAO peak measurements, while dotted 
lines (cross denotes best-fit point) are derived using SNe Ia and BAO data only.
Numerical noise is responsible for the jaggedness of some parts of contours. The
best-fit parameters in both cases are: $\Omega_{\rm m}=0.24$, $\alpha=0.30$ with $\chi^2=326$ for $313$ degrees of freedom (solid lines) and $\chi^2=321$ for $307$ degrees of freedom (dotted lines). Note the different axes scales compared 
to Figure\ 3.}
\end{figure}

\begin{figure}
\includegraphics[clip, trim=2mm 0 1mm 0, scale=1.3]{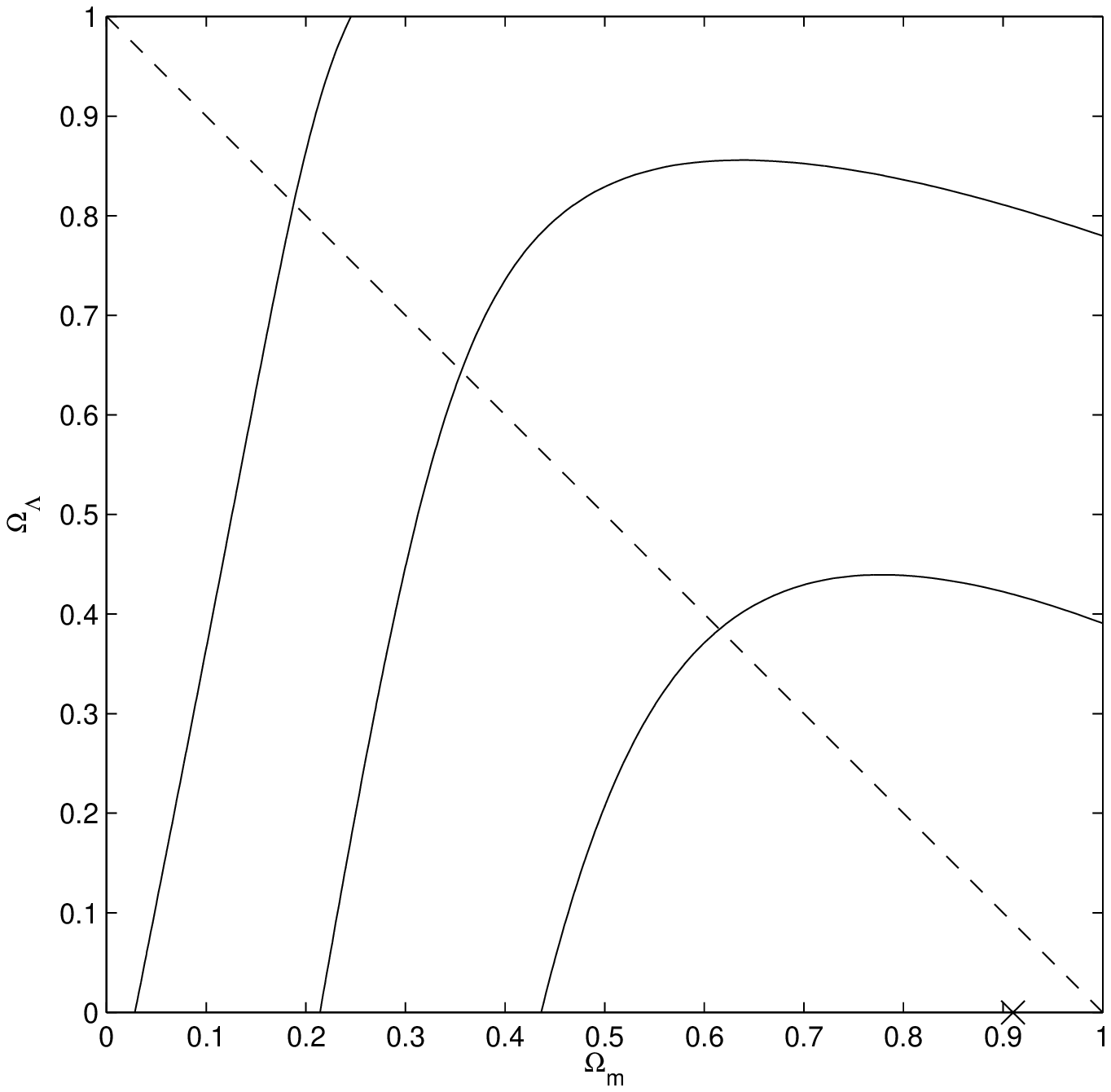}
\caption{1$\sigma$, 2$\sigma$, and 3$\sigma$ confidence level contours for the $\Lambda$CDM 
model from the GRB data, derived using the method of \citet{schaefer07}. The cross indicates best-fit parameter values $\Omega_{\rm m}=0.91$, $\Omega_\Lambda=0.0$ with $\chi^2=77.86$ for $67$ degrees of freedom. The dashed line demarcates spatially flat $\Lambda$CDM models.}
\end{figure}

\begin{figure}
\includegraphics[clip, trim=2mm 0 1mm 0, scale=1.3]{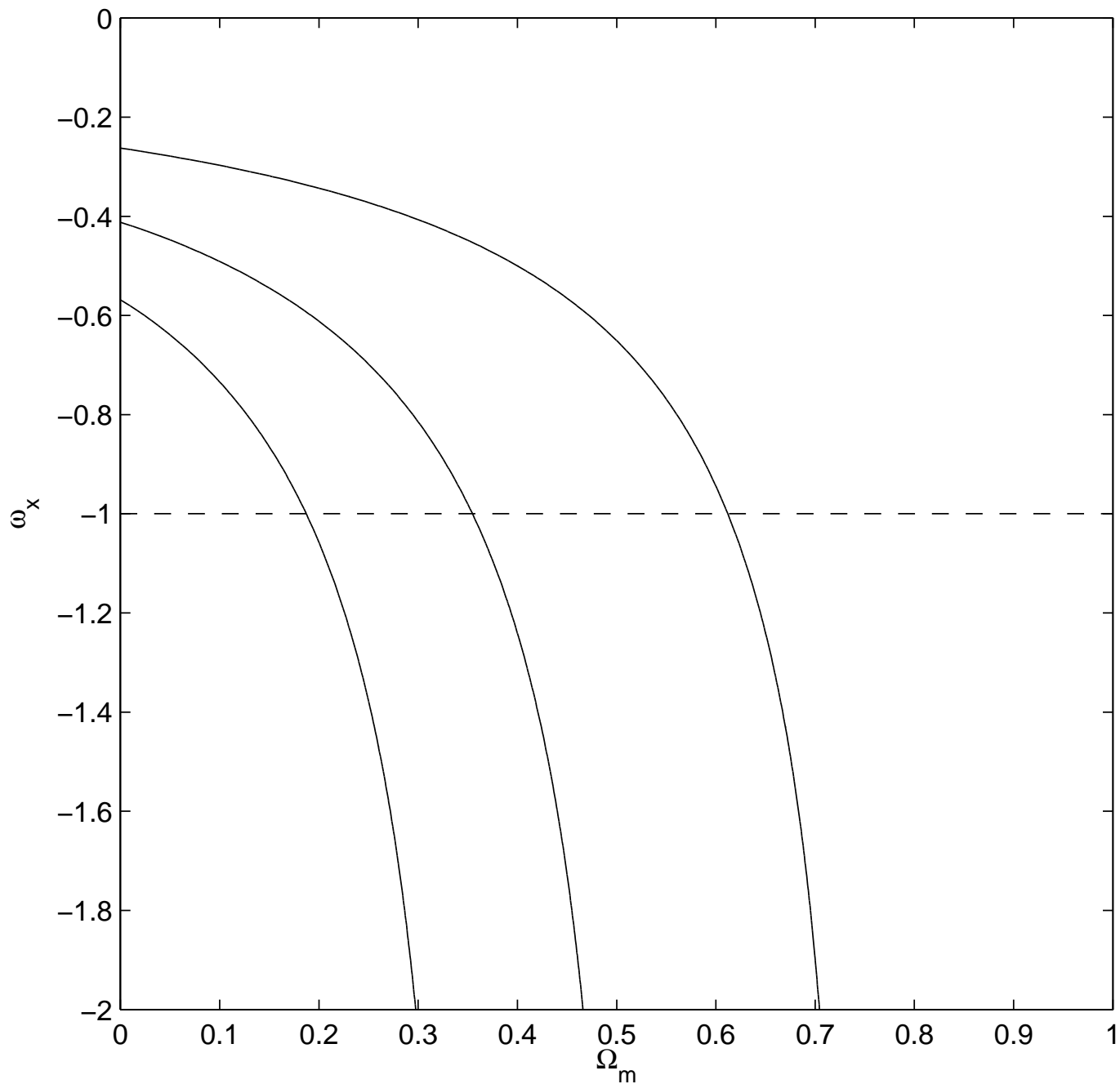}
\caption{1$\sigma$, 2$\sigma$, and 3$\sigma$ confidence level contours for the XCDM model from 
the GRB data, derived using the method of \citet{schaefer07}. The best fit is 
achieved on the line $\omega_{\rm x}=0.00$, which corresponds to the 
spatially flat matter-dominated Universe, $\chi^2=77.8$ for $67$ degrees of freedom. The dashed line demarcates spatially 
flat $\Lambda$CDM models.}
\end{figure}

\begin{figure}
\includegraphics[clip, trim=2mm 0 1mm 0, scale=1.3]{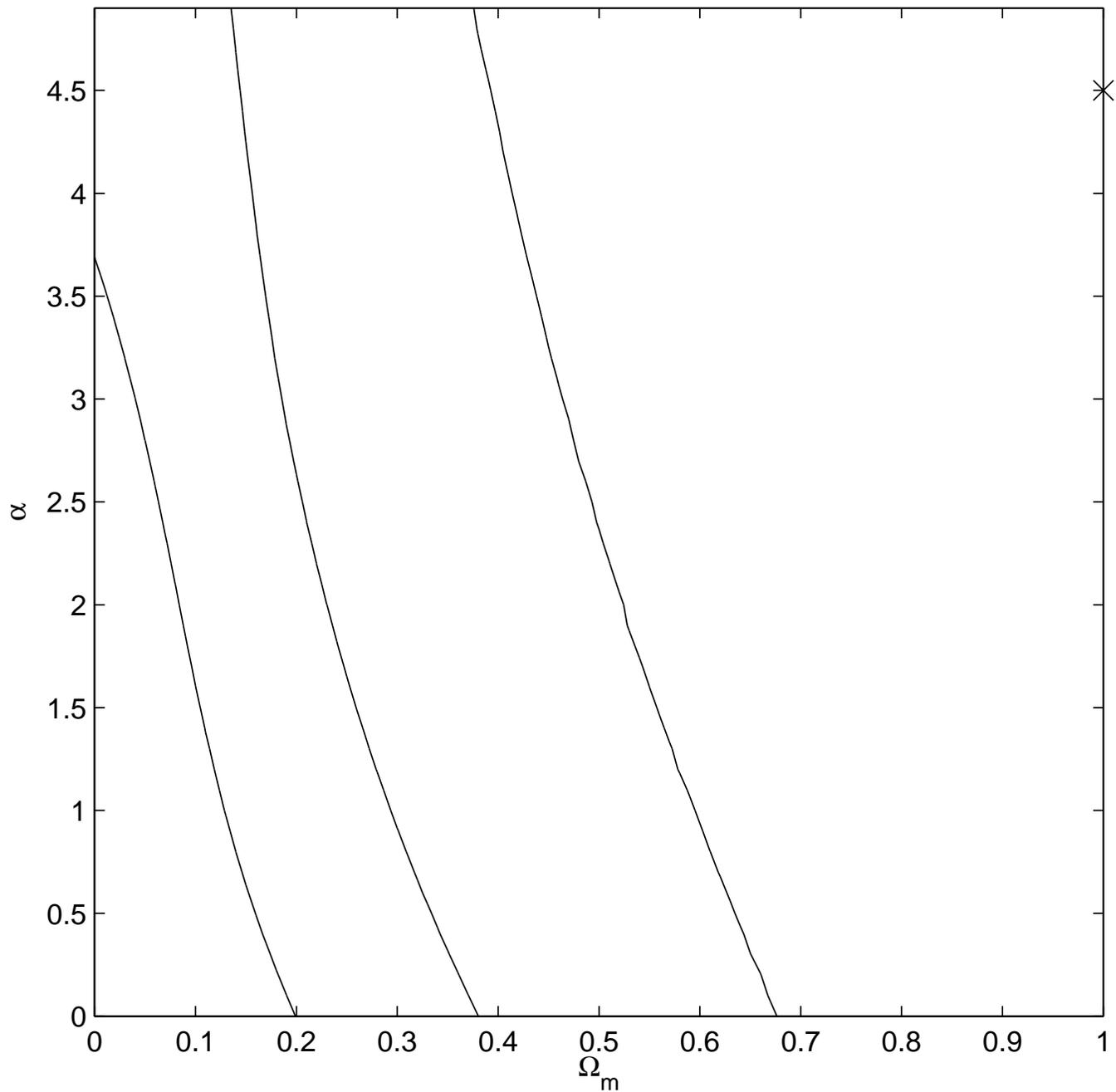}
\caption{1$\sigma$, 2$\sigma$, and 3$\sigma$ confidence level contours for the $\phi$CDM model 
from the GRB data, derived using the method of \citet{schaefer07}. The cross indicates best-fit parameter values $\Omega_{\rm m}=1.0$, $\alpha=4.5$ with $\chi^2=77.8$ for $67$ degrees of freedom. The $\alpha=0$ horizontal axis corresponds to the spatially flat $\Lambda$CDM case.}
\end{figure}

\begin{figure}
\includegraphics[clip,trim=2mm 0 1mm 0, scale=1.3]{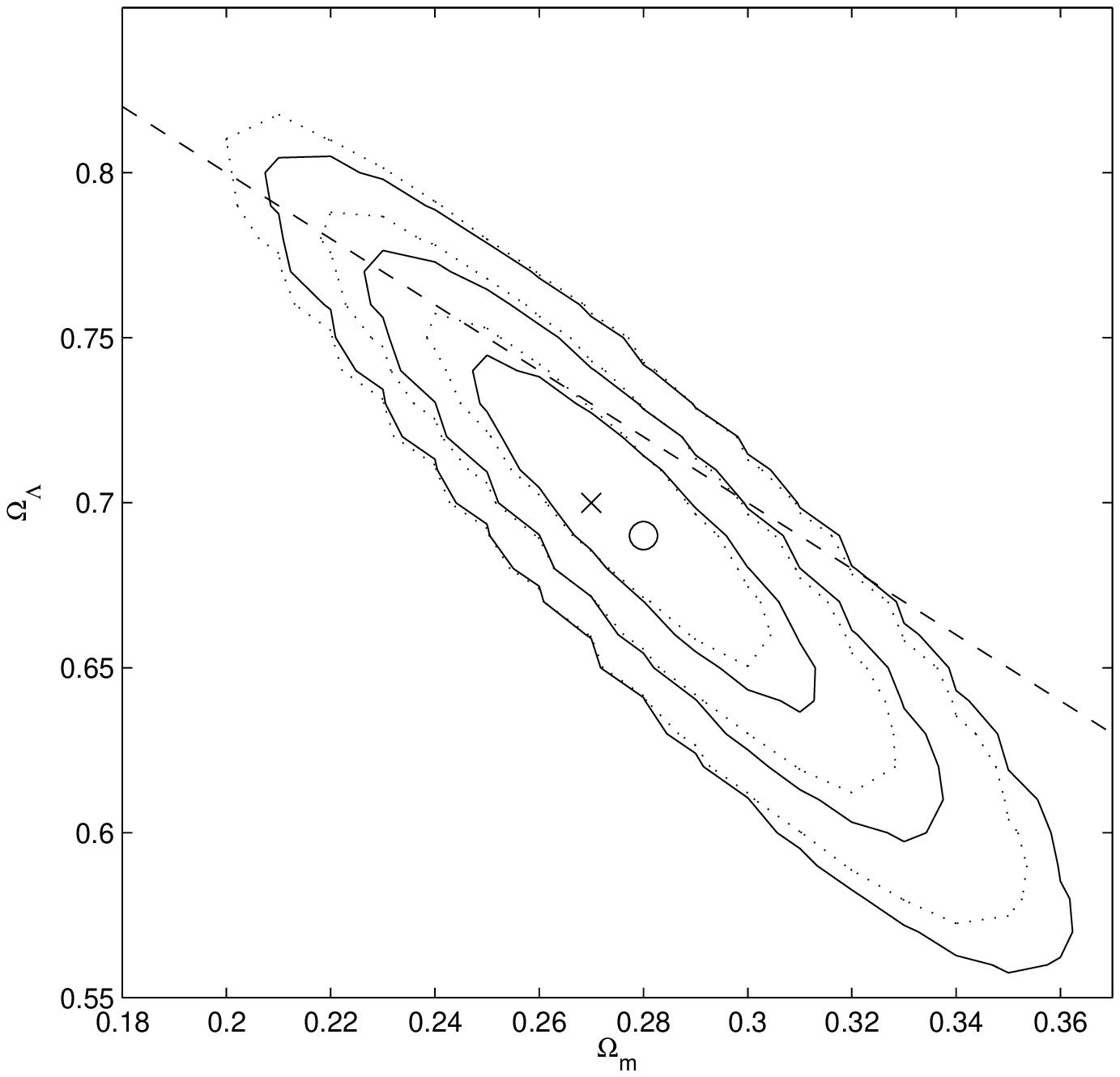}
\caption{1$\sigma$, 2$\sigma$, and 3$\sigma$ confidence level contours for the $\Lambda$CDM 
model (dashed line demarcates spatially flat models). Solid lines (circle 
denotes the best-fit point) are derived using the GRB data (method of 
\citep{schaefer07}), SNe Ia Union data, and BAO peak measurements, while dotted 
lines (cross denotes the best-fit point) are derived using SNe Ia and BAO data 
only. Numerical noise is responsible for the jaggedness of some parts of the
contours. The best-fit parameters are $\Omega_{\rm m}=0.27$,
$\Omega_\Lambda=0.7$ with $\chi^2=321$ for $307$ degrees of freedom (dotted
lines) and $\Omega_{\rm m}=0.28$, $\Omega_\Lambda=0.69$ with $\chi^2=401$ for
$376$ degrees of freedom (solid lines). Note the different axes scales compared
to Figure\ 7.}
\end{figure}

\begin{figure}
\includegraphics[clip, trim=2mm 0 1mm 0, scale=1.3]{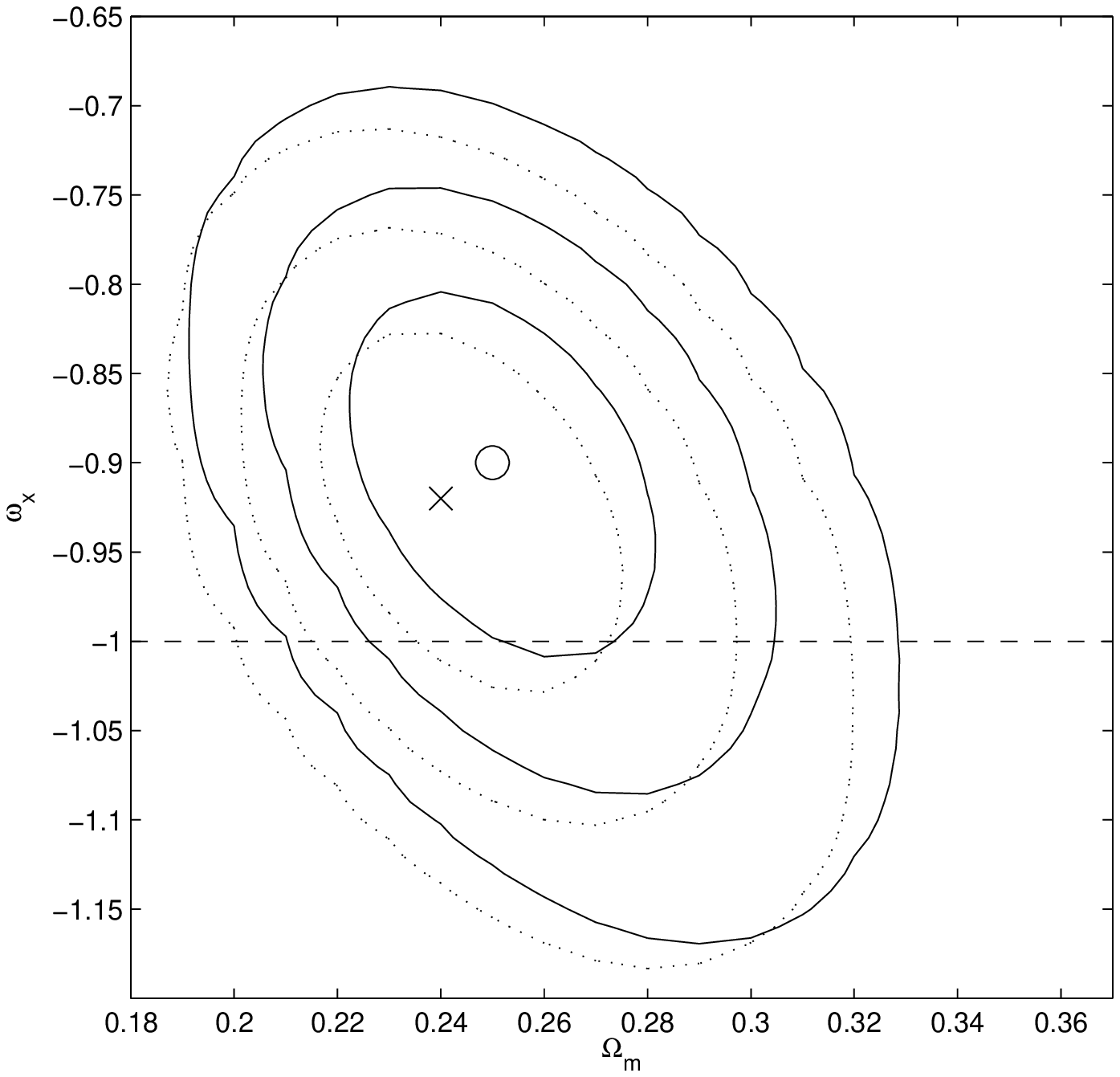}
\caption{1$\sigma$, 2$\sigma$, and 3$\sigma$ confidence level contours for the XCDM model. 
Solid lines (circle denotes the best-fit point) are derived using GRB data 
(method of \citep{schaefer07}), SNe Ia Union data, and BAO peak measurements,
while dotted lines (cross denotes the best-fit point) are derived using 
SNe Ia and BAO peak data only. The best-fit parameter values are, for solid
contours (circle) -- $\Omega_{\rm m}=0.26$, $\omega_{\rm x}=-0.90$ with
$\chi^2=401$ for $376$ degrees of freedom, and for dotted contours (cross)  --
$\Omega_{\rm m}=0.25$, $\omega_{\rm x}=-0.91$ with $\chi^2=322$ for $307$
degrees of freedom. Note the different axes scales compared to Figure\ 8.}
\end{figure}

\begin{figure}
\includegraphics[clip, trim=2mm 0 1mm 0, scale=1.3]{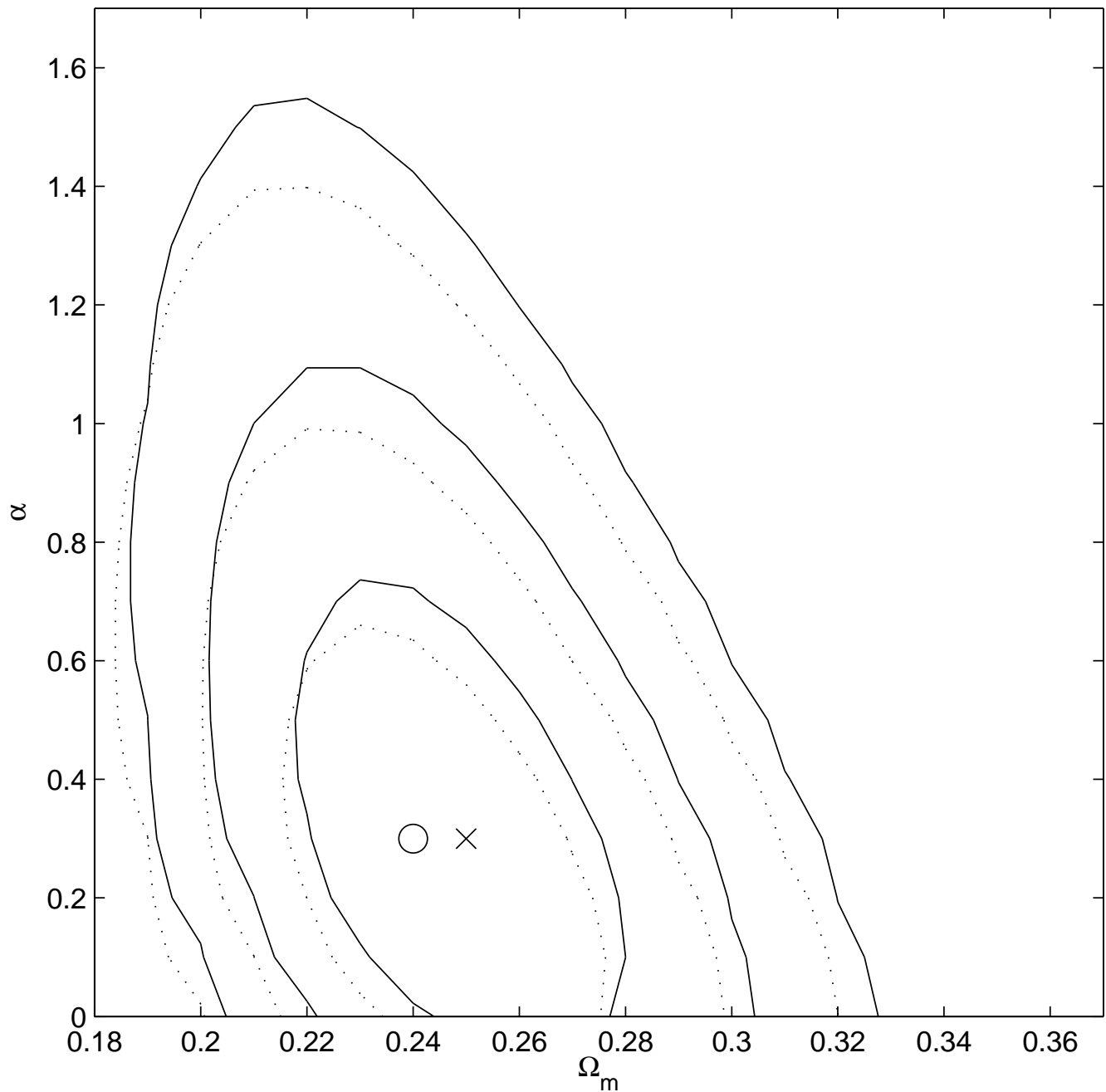}
\caption{1$\sigma$, 2$\sigma$, and 3$\sigma$ confidence level contours for the $\phi$CDM model.
Solid lines (circle denotes best-fit point) are derived using GRB data (method 
of \citep{schaefer07}), SNe Ia Union data, and BAO peak measurements, while dotted
lines (cross denotes best-fit point) are derived using SNe Ia and BAO data only.
The best-fit parameters are $\Omega_{\rm m}=0.24$, $\alpha=0.30$ with
$\chi^2=401$ for $376$ degrees of freedom (solid lines) and $\Omega_{\rm
m}=0.25$, $\alpha=0.30$ with $\chi^2=321$ for $307$ degrees of freesom (dotted
  lines). Note different axes scales compared to Figure\ 9.}
\end{figure}


\begin{thebibliography}{}

\bibitem[Alam et al(2009)]{alam09}
  Alam, U., Sahni, V., \& Starobinsky, A.~A.~2009, \apj, 704, 1086 

\bibitem[Allen et al.(2008)]{allen08}
  Allen, S.~W., et al.~2008, \mnras, 383, 879

\bibitem[Aluri et al.(2009)]{aluri09}
  Aluri, P.~K., Jain, P., \& Singh, N.~K.~2009, Mod. Phys. Lett. A, 24, 1583

\bibitem[Amati et al.(2008)]{amati08}
  Amati, L., et al.~2008, \mnras, 391, 577

\bibitem[Andrianov et al.(2010)]{andrianov09}
  Andrianov, A.~A., Cannata, F., Kamenshchik, A.~Y., \& Regoli, D.~2010, Int. J.
  Mod. Phys. D., 19, 97
   
\bibitem[Arun et al.(2009)]{arun09}
  Arun, K.~G., et al.~2009, Class. Quant. Grav. 26, 094021

\bibitem[Bamba et al.(2009)]{bamba09}
  Bamba, K., Geng, C.-Q., Nojiri, S., \& Odintsov, S.~D.~2009, \prd, 79, 083014

\bibitem[Basilakos \& Perivolaropoulos(2008)]{basilakosperivolaropoulos08}
  Basilakos, S., \& Perivolaropoulos, L.~2008, \mnras, 391, 411
  
\bibitem[Bertolami \& Silva(2006)]{bertolumi06}
  Bertolami, O., \& Silva, P.~T.~2006, \mnras, 365, 1149

\bibitem[Bili\'{c} et al.(2009)]{bilic09}
  Bili\'{c}, N., Tupper, G.~A., \& Viollier, R.~D.~2009, \prd, 80, 023515

\bibitem[Caldwell \& Kamionkowski(2009)]{caldwell09}
  Caldwell, R.~R., \& Kamionkowski, M.~2009, Ann. Rev. Nucl. Part. Sci., 59, 397

\bibitem[Capozziello et al.(2008)]{capozziello08}
  Capozziello, S., Cianci, R., Stornaiolo, C., \& Vignolo, S.~2008, Phys. Scr., 78, 065010
  
\bibitem[Capozziello \& Izzo(2008)]{capozzielloizzo08}
  Capozziello, S., \& Izzo, L.~2008, \aap, 490, 31
  
\bibitem[Casarini et al.(2009)]{casarini09}
  Casarini, L., Macci\`{o}, A.~V., \& Bonometto, S.~A.~2009, \jcap, 0903, 014
  
\bibitem[Chae et al.(2004)]{chae04}
  Chae, K.-H., Chen, G., Ratra, B., \& Lee, D.-W.\ 2004, ApJ, 607, L71

\bibitem[Chen  \& Ratra(2003)]{chen03a}
  Chen, G., \& Ratra, B.~2003, ApJ, 582, 586

\bibitem[Chen \& Ratra(2004)]{chen04}
  Chen, G., \& Ratra, B.~2004,  ApJ, 612, L1

\bibitem[Chongchitnan(2009)]{chong09}
  Chongchitnan, S.~2009, \prd, 79, 043522

\bibitem[Coc et al(2009)]{coc09}
  Coc, A., Olive, K.~A., Uzan, J.-P., \& Vangioni, E.~2009, \prd, 79, 103512

\bibitem[Cole et al.(2005)]{cole05}
  Cole, S., et al.~2005, \mnras, 362, 505

\bibitem[Cunha(2009)]{cunha09}
  Cunha, J.~V.~2009, \prd, 79, 047301

\bibitem[Daly et al.(2009)]{daly09}
  Daly, R.~A., et al.~2009, \apj, 691, 1058

\bibitem[Dent et al.(2009)]{dent09}
  Dent, T., Stern, S., \& Wetterich, C.~2009, \jcap, 0901, 038

\bibitem[Dev et al.(2008)]{dev08}
  Dev, A., Jain, D., \& Lohiya, D.~2008, arXiv:0804.3491 [astro-ph]
  
\bibitem[Di Girolamo et al.(2005)]{digirolamo05}
  Di Girolamo, T., Catena, R., Vietri, M., \& Di Sciascio, G.~2005, \jcap, 0504, 008

\bibitem[Dunkley et al.(2009)]{dunkley09}
  Dunkley, J., et al.~2009, \apjs, 180, 306
  
\bibitem[Dutta \& Scherrer(2009)]{dutta09}
  Dutta, S., \& Scherrer, R.~J.~2008, \prd, 78, 123525

\bibitem[Eisenstein et al.(2005)]{eisensteinetal}
  Eisenstein, D.~J., et al.~2005, \apj, 633, 560

\bibitem[Ettori et al.(2009)]{ettori09}
  Ettori, S., et al.~2009, \aa, 501, 61

\bibitem[Feng(2009)]{feng09}
  Feng. C.-J.~2009, Phys. Lett. B, 672, 94

\bibitem[Fernandez-Martinez \& Verde(2008)]{fernandezmartinez08}
  Fernandez-Martinez, E., \& Verde, L.~2008, \jcap, 0808, 023
  
\bibitem[Firmani et al.(2005)]{firmani05}
  Firmani, C., Ghisellini, G., Ghirlanda, G., \& Avila-Reese, V., 2005, \mnras, 360, L1
  
\bibitem[Francis et al.(2008)]{francis09}
  Francis, M.~J., Lewis, C.~F., \& Linde, E.~V.~2008, \mnras, 393, L31
  
\bibitem[Friedman \& Bloom(2005)]{friedman05}
  Friedman, A.~S., \& Bloom, J.~S.~2005, \apj, 627, 1
  
\bibitem[Frieman(2009)]{frieman09}
  Frieman, J.~A.~2009, arXiv:0904.1832 [astro-ph]
 
\bibitem[Frieman et al.(2008)]{frieman08}
  Frieman, J.~A., Turner, M.~S., \& Huterer, D.~2008, \araa, 46, 385

\bibitem[Ghirlanda et al.(2004)]{ghirlanda04}
  Ghirlanda, G., Ghisellini, G., Lazzati, D., \& Firmani, C.~2004, \apj, 613, L13
 
\bibitem[Gong et al.(2009)]{gong09}
  Gong, Y., Zhang, T.-J., Lan, T., \& Chen, X.-L.~2009, arXiv:0810.3572 [astro-ph]

\bibitem[Grande et al.(2009)]{grande09}
  Grande, J., Pelinson, A., \& Sol\`{a}, J.~2009, \prd, 79, 043006

\bibitem[Grossi \& Springel(2009)]{grossi09}
  Grossi, M., \& Springel, V.~2009, arXiv:0809.3404 [astro-ph]

\bibitem[Harko(2008)]{harko08}
  Harko, T.~2008, Phys. Lett. B, 669, 376
  
\bibitem[Hellwing \& Juszkiewicz(2009)]{hellwing09}
  Hellwing, W.~A. \& Juszkiewicz, R.~2009, \prd, 80, 083522

\bibitem[Hicken et al.(2009)]{hicken09}
  Hicken, M. et al.~2009, \apj, 700, 1097

\bibitem[Isobe et al.(1990)]{isobe90}
  Isobe, T., Feigelson, E.~D., Akritas, M.~G., \& Babu, G.~J.~1990, \apj, 364, 104

\bibitem[Jamil(2009)]{jamil09}
  Jamil, M.~2009, arXiv:0810.2896 [gr-qc]

\bibitem[Kilbinger et al.(2009)]{kilbinger09}
  Kilbinger, M., et al.~2009, \aap, 497, 677

\bibitem[Kodama et al.(2008)]{kodama08}
  Kodama, Y., et al.~2008, \mnras, 391, L1
  
\bibitem[Komatsu et al.(2009)]{komatsu09}
  Komatsu, E., et al.~2009, \apjs, 180, 330

\bibitem[Kowalski et al.(2008)]{kowalski09}
  Kowalski, M., et al.~2008, \apj, 686, 749

\bibitem[Lamb \& Reichart(2000)]{lamb00}
  Lamb, D.~Q. \& Reichart, D.~E.~2000, \apjl, 536, 1

\bibitem[Lamb et al.(2005)]{lamb05}
  Lamb, D.~Q., et al.~2005, arXiv:astro-ph/0507362

\bibitem[La Vacca et al.(2009)]{lavacca09}
  La Vacca, G., Bonometto, S.~A., \& Colombo, L.~P.~L.~2009, New Astron. 14, 435

\bibitem[Liang et al.(2008)]{liangetal08}
  Liang, N., Xiao, W.~K., Liu, Y., \& Zhang, S.~N.~2008, \apj, 685, 354

\bibitem[Liang \& Zhang(2008)]{liang08}
  Liang, N. \& Zhang, L.~N.~2008, AIP Conf. Proc., 1065, 367
  
\bibitem[Lin et al.(2008)]{lin08}
  Lin, H., Zhang, T.-J., \& Yuan, Q.~2008, Mod. Phys. Lett. A, 24, 1699

\bibitem[M\'{e}sz\'{a}ros(2006)]{meszaros06}
  M\'{e}sz\'{a}ros, P.~2006, Rep. Prog. Phys., 69, 2259

\bibitem[Mortonson et al.(2009)]{mortonson09}
  Mortonson, M.~J., Hu, W., \& Huterer, D.~2009, \prd, 79, 023004
  
\bibitem[M\"{o}rtsell \& Sollerman(2005)]{mortsell05}
  M\"{o}rtsell, E., \& Sollerman, J.~2005, \jcap, 0506, 009

\bibitem[Mosquera Cuesta et al.(2008)]{mosquera08}
  Mosquera Cuesta, H.~J., Dumet, M.,~H., \& Furlanetto, C.~2008, \jcap, 0807, 004

\bibitem[Nemiroff(2000)]{nemiroff00}
  Nemiroff, R.~J.~2000, \apj, 544, 805

\bibitem[Nesseris(2009)]{nesseris09}
  Nesseris, S.~2009, \prd, 79, 044015

\bibitem[Page et al.(2003)]{page03}
  Page, L., et al.~2003, ApJS, 148, 233

\bibitem[Peebles(1984)]{pee84}
  Peebles, P.~J.~E.~1984, \apj, 284, 439

\bibitem[Peebles \& Ratra(1988)]{peebles88}
  Peebles, P.~J.~E., \&\ Ratra, B.~1988, \apj, 325, L17

\bibitem[Peebles \& Ratra(2003)]{peebles03}
  Peebles, P.~J.~E., \&\ Ratra, B.~2003, Rev.~Mod.~Phys., 75, 559

\bibitem[Percival et al.(2007)]{percivaletal07a}
  Percival, W.~J., et al.~2007, \mnras, 381, 1053 

\bibitem[Perivolaropoulos(2009)]{perivolaropoulos09}
  Perivolaropoulos, L.~2009, arXiv:0811.4684 [astro-ph]
  
\bibitem[Perivolaropoulos \& Shafieloo(2009)]{perivolaropoulosshafieloo09}
  Perivolaropoulos, L., \& Shafieloo, A.~2009, \prd, 79, 123502

\bibitem[Podariu et al.(2003)]{podariu03}
  Podariu, S., Daly, R.\ A., Mory, M., \& Ratra, B.\ 2003 ApJ, 584, 577

\bibitem[Podariu et al.(2001a)]{podariu01a}
  Podariu, S., Nugent, P., \& Ratra, B.~2001a, \apj, 553, 39

\bibitem[Podariu et al.(2001b)]{podariu01b}
  Podariu, S., Souradeep, T., Gott, J.~R., Ratra, B., \& Vogeley, M.~S.~2001b, ApJ, 559, 9

\bibitem[Primack(2009)]{primack09}
  Primack, J.~R.~2009, arXiv:0902.2506 [astro-ph.CO]

\bibitem[Qi et al(2008)]{qi09}
  Qi, S., Wang, F.-Y., \& Lu, T.~2008, \aa, 487, 853

\bibitem[Ratra(1991)]{ratra91}
  Ratra, B.~1991, \prd, 43, 3802

\bibitem[Ratra \& Peebles(1988)]{ratra88}
  Ratra, B., \& Peebles, P.~J.~E.~1988, \prd, 37, 3406

\bibitem[Ratra \& Vogeley(2008)]{ratra08}
  Ratra, B., \& Vogeley, M.~S.~2008, \pasp, 120, 235

\bibitem[Sahni et al.(2008)]{sahni08}
  Sahni, V., Shafieloo, A., \& Starobinsky, A.~A.~2008, \prd, 78, 103502
 
\bibitem[Sami(2009)]{sami09}
  Sami, M.~2009, Curr. Sci., 97 887
   
\bibitem[Samushia et al.(2007)]{samushia07}
  Samushia, L., Chen, G., \& Ratra, B.~2007, arXiv:0706.1963 [astro-ph]

\bibitem[Samushia \& Ratra(2006)]{samushia06}
  Samushia, L., \& Ratra, B.~2006, \apj, 650, L5

\bibitem[Samushia \& Ratra(2008)]{samushia08}
  Samushia, L., \& Ratra, B.~2008, \apj, 680, L1

\bibitem[Samushia \& Ratra(2009a)]{samushia09a}
  Samushia, L., \& Ratra, B.~2009a, \apj, 703, 1904
  
\bibitem[Samushia \& Ratra(2009b)]{samushia09b}
  Samushia, L., \& Ratra, B.~2009b, \apj, 701, 1373
  
\bibitem[Santos \& Lima(2008)]{santos08}
  Santos, R.~C., \& Lima, J.~A.~S.~2008, \prd, 77, 083505

\bibitem[Schaefer(2007)]{schaefer07}
  Schaefer, B.~E.~2007, \apj, 660, 16

\bibitem[Sen \& Devi(2010)]{sen09}
  Sen, A.~A., \& Devi, N.~C.~2010, Gen. Rel. Grav., 42, 821

\bibitem[Setare \& Saridakis(2009)]{setare09}
  Setare, M.~R., \& Saridakis, E.~N.~2009, Phys. Lett. B, 671, 331
  
\bibitem[Shaposhnikov \& Zenh\"{a}usern(2009)]{shaposhnikov09}
  Shaposhnikov, M., \& Zenh\"{a}usern, D.~2009, Phys. Lett. B, 671, 87

\bibitem[Silvestri \& Trodden(2009)]{silvestri09}
  Silvestri, A., \& Trodden, M.~2009, Rep. Prog. Phys., 72, 096901

\bibitem[Tanvir et al.(2009)]{tanvir09}
  Tanvir, N.~R., et al.~2009, Nature, 461, 1254 

\bibitem[Thomas et al.(2009)]{thomas09}
  Thomas, S.~A., Abdalla, F.~B., \& Weller, J.~2009, \mnras, 395, 197
  
\bibitem[Tsutsui et al.(2009)]{tsutsui09}
  Tsutsui, R., et al.~2009, \mnras, 394, L31
  
\bibitem[Wang \& Zhang(2008)]{wangzhang08}
  Wang, S., \& Zhang, Y.~2008, Phys. Lett. B, 669, 201
  
\bibitem[Wang et al.(2009)]{wang09}
  Wang X., et al.~2009, \mnras, 394, 1775
  
\bibitem[Wang(2008)]{yuwang08}
  Wang, Y.~2008, \prd, 78, 123532

\bibitem[Wei \& Zhang(2008)]{wei08}
  Wei, H., \& Zhang, S.~N.~2008, Eur. Phys. J. C., 63, 139

\bibitem[Wilson et al.(2006)]{wilson06}
  Wilson, K. M., Chen, G., \& Ratra, B.~2006, Mod. Phys. Lett. A, 21, 2197

\bibitem[Xu et al.(2005)]{xu05}
  Xu, D., Dai, Z.~G., \& Liang, E.~W.~2005, \apj, 633, 603

\bibitem[Yashar et al.(2009)]{yashar09}
  Yashar, M., et al.~2009, \prd, 79, 103004

\end{thebibliography}
\end{document}